\documentclass[%
 reprint,
 amsmath,amssymb,
 aps,
]{revtex4-1}

\usepackage{graphicx}
\usepackage{dcolumn}
\usepackage{bm}

\def\half{{\textstyle{1\over 2}}}
\def\frac#1#2{{\textstyle{{#1}\over {#2}}}}

\def\sb{\overline{s}{}}

\def\re{{\rm Re}~}
\def\im{{\rm Im}~}

\newcommand{\beq}{\begin{equation}}
\newcommand{\eeq}{\end{equation}}
\newcommand{\bea}{\begin{eqnarray}}
\newcommand{\eea}{\end{eqnarray}}

\newcommand{\rf}[1]{(\ref{#1})}

\begin{document}


\title{Measuring the Speed of Gravitational Waves from the First and Second Observing Run of Advanced LIGO and Advanced Virgo}

\author{Xiaoshu Liu$^1$}
\email{xiaoshu@uwm.edu}
\author{Vincent F.\ He$^2$}
\author{Timothy M.\ Mikulski$^2$}
\author{Daria Palenova$^2$}
\author{Claire E.\ Williams$^2$}
\author{Jolien Creighton$^1$}
\email{jolien@uwm.edu}
\author{Jay D.\ Tasson$^2$}
 \email{jtasson@carleton.edu}
\affiliation{
 $^1$Department of Physics, University of Wisconsin-Milwaukee, Milwaukee, WI 53201, USA\\
$^2$Department of Physics and Astronomy, Carleton College,
Northfield, MN 55057, USA
}

\begin{abstract}
The speed of gravitational waves for a single observation can be measured by the time delay among gravitational-wave detectors with Bayesian inference. Then multiple measurements can be combined to produce a more accurate result. From the near simultaneous detection of gravitational waves and gamma rays originating from GW170817/GRB 170817A, the speed of gravitational wave signal was found to be the same as the the speed of the gamma rays to approximately one part in $10^{15}$.  Here we present a different method of measuring the speed of gravitational waves, not based on an associated electromagnetic signal but instead by the measured transit time across a geographically separated network of detectors.  While this method is far less precise, it provides an independent measurement of the speed of gravitational waves. For GW170817 a binary neutron star inspiral observed by Advanced LIGO and Advanced Virgo, by fixing sky localization of the source at the electromagnetic counterpart the speed of gravitational waves is constrained to 90\% confidence interval (0.97c, 1.02c), where c is the speed of light in a vacuum. By combing ten BBH events and the BNS event from the first and second observing run of Advanced LIGO and Advanced Virgo, the 90\% confidence interval is narrowed down to (0.97c, 1.01c). The accurate measurement of the speed of gravitational waves allows us to test the general theory of relativity.
We further interpret these results
within the test framework provided by the gravitational Standard-Model Extension (SME).
In doing so,
we obtain simultaneous constraints on 4 of the 9 nonbirefringent, nondispersive coefficients for Lorentz violation in the gravity sector of the SME
and place limits on the anisotropy of the speed of gravity.

\end{abstract}

\pacs{Valid PACS appear here}
\maketitle


\section{\label{sec:intro}Introduction}

The first gravitational wave (GW) detection, GW150914 \cite{LVC:GW150914}, was observed from a binary black hole (BBH) merger during the first observing run(O1) of Advanced LIGO \cite{TheLIGOScientific:2014jea} from September 12th, 2015 to January 19th, 2016. Later in O1, two BBH mergers GW151012 \cite{LVC:O1} and GW151226 \cite{LVC:GW151226} were also detected by the two Advanced LIGO detectors. The second observing run (O2) of the Advanced LIGO took place from November 30th, 2016 to August 25th, 2017.  In O2 three BBH mergers GW170104 \cite{LVC:GW170104}, GW170608 \cite{LVC:GW170608} and GW170823 \cite{LVC:Catalog} were detected by the two Advanced LIGO detectors.  With the Advanced Virgo \cite{TheVirgo:2014hva} detector joining in later O2, four more BBH mergers GW170729 \cite{LVC:Catalog}, GW170809 \cite{LVC:Catalog}, GW170814 \cite{LVC:GW170814} , GW170818 \cite{LVC:Catalog}  and one binary neutron star (BNS) inspiral GW170817 \cite{LVC:GW170817} were observed by the three-detector network \cite{Aasi:2013wya}.    

General Relativity predicts that the speed of gravitational waves is equal to the speed of light in a vacuum. The GW seen by the Advanced LIGO and Advanced Virgo detectors can be used to test the theory of general relativity. The first measurement of the speed of gravitational waves using time delay among the GW detectors was suggested by Cornish \textit{et al} \cite{Cornish:2017sgw}. By applying the Bayesian method the speed of gravitational waves is constrained to 90\% confidence interval between 0.55$c$ and 1.42$c$ with GW150914, GW151226 and GW170104 \cite{Cornish:2017sgw}. 

Subsequent to Cornish \textit{et al} \cite{Cornish:2017sgw}, a more precise measurement of the speed of gravitational waves was facilitated by the measurement of the time delay between GW and electromagnetic observations of the same astrophysical source. On August 17, 2017, a binary neutron star inspiral GW170817 was observed by the Advanced LIGO and Advanced Virgo detectors, (1.74 $\pm$ 0.05)s later the Gamma-ray burst(GRB) was observed independently by Fermi Gamma-ray Laboratory. By using the lower bound of luminosity distance obtained from the GW signal, the time delay between the GW and GRB, and some astrophysical assumptions, the speed of gravitational waves($v_g$) was constrained to $-3 \times 10^{-15}c < v_g-c < +7 \times 10^{-16}c$ \cite{LVC:2017gw-grb}.

Using the time delay between GW and GRB requires assuming 
the time difference of emission of the gamma rays relative to the peak of the gravitational waveform. For extreme models, this difference could be $\sim 1000s$ \cite{{2015APS..APRE14004C}, {2015ApJ...802...95R}} , which is much larger than the $10s$ lag adopted in Ref. \cite{LVC:2017gw-grb}, and emission of the gamma rays could even lead the merger \cite{2012PhRvL.108a1102T}, leading to a 2 order of magnitude increase in the range of the constrained speed on either side. While comparing the arrival time of the gravitational waves to the arrival time of the gamma rays requires various model assumptions, not present in the direct method, the precision of this method nevertheless vastly exceeds what could ever be obtained with the direct method presented here.  Disagreement between the direct method and the electromagnetic counterpart method would be nearly inexplicable.  Not surprisingly, we find no such disagreement.

In this paper, we employ an approach similar to that used in Ref. \cite{Cornish:2017sgw}, to make a local measurement of the speed of gravity based on the difference in arrival time across a network of GW detectors for pure GW observations made during O1 and O2.  We consider both measurements of the speed of gravitational waves from individual events and then demonstrate how the accuracy can be improved by combining measurements from multiple GW observations.  In Sec.\ \ref{sec:methods}, we discuss our methods and in Sec.\ \ref{sec:results} we present the speed of gravity results.
Finally, in Sec.\ \ref{sec:lorentz}, we use a subset of the individual speed of gravity results
to obtain constraints on local Lorentz violation in the context of the effective-field-theory-based test framework provided by the gravitational Standard-Model Extension (SME) \cite{[{For an annually updated review of observational and experimental results, see }] data,[{For early foundational work on the SME, see }]ck,[{For foundational gravity-sector work, see }]akgrav,*lvpn}, within which a number of recent theoretical \cite{kmgw, mewes2019, *xu2019} and experimental \cite{LVC:2017gw-grb,shao2020} studies of GWs have been performed.  While the results achieved here are much weaker than those attained via multimessenger astronomy in Ref.\ \cite{LVC:2017gw-grb}, the analysis presented here offers several novel features.  In Ref.\ \cite{LVC:2017gw-grb}, constraints on SME coefficients were attained using a maximum-reach approach \cite{flowers2016}, effectively constraining a series of 9 models having one parameter each.  Here, we attain simultaneous constraints on multiple coefficients for Lorentz violation using direct observations of the speed of gravity for the first time.  In addition, the approach is quite different from both that of Ref. \cite{LVC:2017gw-grb} and those of earlier works \cite{data} and hence is subject to a different set of assumptions. For example, the current approach is free of the astrophysical-modeling assumptions used in Ref. \cite{LVC:2017gw-grb}. This also provides the first direct limits on direction-dependent GW speeds.

\section{\label{sec:methods}Methods}

\subsection{\label{sec:level2}Measuring the speed of gravitational waves with a single GW event}
The standard parameter estimation based on GW data from multiple detectors imposes the constraint that the signal propagation across the network is at the speed of light \citep{lalsuite}. It first generates a random time at Earth center within a small time window ($\pm 0.1s$) of an arrival time reported by a search pipeline, and then $v_g$ is used to compute the corresponding time at each detector in order to generate waveform templates. In this work, however, we remove this constraint, allowing $v_g$ to be a parameter in order to be estimated along with all other signal parameters.

GW data $d_i$ collected at detector i, can be decomposed into pure GW signal $h_i(t)$ plus random noise $n_i(t)$:
\begin{equation}
d_i(t) = h_i(t) + n_i(t)
\end{equation} 
the posterior distribution of a set of parameters $\vec{\theta}$ can be obtained via Bayes' theorem:
\begin{align} 
p( \vec{\theta}|d_1,d_2,...) &= \frac{p(\vec{\theta})p(d_1,d_2,...|\vec{\theta})}{p(d_1,d_2,...)}\\
&\propto p(\vec{\theta})p(d_1,d_2,...|\vec{\theta})
\end{align}
Where $p(\vec{\theta})$ is the prior distribution which reflects what we know about $\vec{\theta}$  before the measurement. $p(d_1,d_2,...)=\int p(\vec{\theta})p(d_1,d_2,...|\vec{\theta}) d\vec{\theta}$ is a normalization factor known as evidence which is independent of $\vec{\theta}$ and it is useful for model selection. Assuming the noise is stationary and Gaussian distributed, the likelihood $p(d_1,d_2,...|\vec{\theta})$ can be written as:
\begin{equation}
p(d_1,d_2,...|\vec{\theta}) \propto \prod_i \exp{[-\int_{-\infty}^\infty \frac{|d_i(f)-h_i(f|\vec{\theta})|^2}{S_i(f)}df]}
\end{equation}
Where $d_i(f)=\int_{-\infty}^\infty d_i(t)e^{-2\pi i ft}dt$ is the Fourier transform of $d_i(t)$. $h_i(f|\vec{\theta})$ is a waveform in the frequency domain. $S_i(f)$ is the noise power spectral density(PSD) which characterizes the sensitivity of the GW detector.

The marginalized posterior of the speed of gravitational waves $v_g$ is obtained by integrating over other parameters:
\begin{equation}
p(v_g|d_1,d_2,...) = \int p( \vec{\theta}|d_1,d_2,...)  d\vec{\theta}^\prime
\end{equation} 
Where $\vec{\theta}^\prime$ is a set of parameters in $\vec{\theta}$ except for $v_g$. Markov Chain Monte Carlo(MCMC) with Metropolis-Hastings algorithm \cite{Metrolis, Hastings, Veitch:2015} is an effective method to sample from multi-dimensional posterior distributions.

\subsection{\label{sec:level2}Combing multiple GW events}
The accuracy of the speed of gravitational waves measurement can be improved by combing multiple GW events. Suppose the GW detectors observed $n$ events with data $d_1, d_2, ..., d_n$,  the posterior of $v_g$ for the joint events can be computed by applying Bayes' theorem and assuming the events are mutually independent:
\begin{equation}
p(v_g|d_1,d_2,...,d_n) \propto \frac{p(v_g|d_1)p(v_g|d_2)...p(v_g|d_n)}{p^{n-1}(v_g)}
\label{eq:combined}
\end{equation}
Where $p(v_g|d_i)$ is the marginalized posterior of $v_g$ for event $i$ and $p(v_g)$ is prior distribution of $v_g$. With uniform prior, the Eq. \ref{eq:combined} is simplified to:
\begin{equation}
p(v_g|d_1,d_2,...,d_n) \propto p(v_g|d_1)p(v_g|d_2)...p(v_g|d_n)
\label{eq:simplified}
\end{equation}
which says that marginalized posterior of $v_g$ for joint events is proportional to the product of marginalized posterior of $v_g$ for a single event.

To estimate how much improvement of the combing measurement, we assume posterior of $v_g$ for $n$ GW events are independent and identical Gaussian distribution, i.e. $p(v_g|d_i) \propto \exp(-(v_g-\mu)^2/(2\sigma^2))$. Then, the combined posterior becomes $p(v_g|d_1, d_2, ..., d_n) \propto \exp(-(v_g-\mu)^2/(2\sigma^2/n))$. Therefore, the combing method is expected to reduce standard deviation of a single measurement by a factor of  $\sqrt{n}$. 

\begin{figure}
\centering
\includegraphics[width=0.5\textwidth]{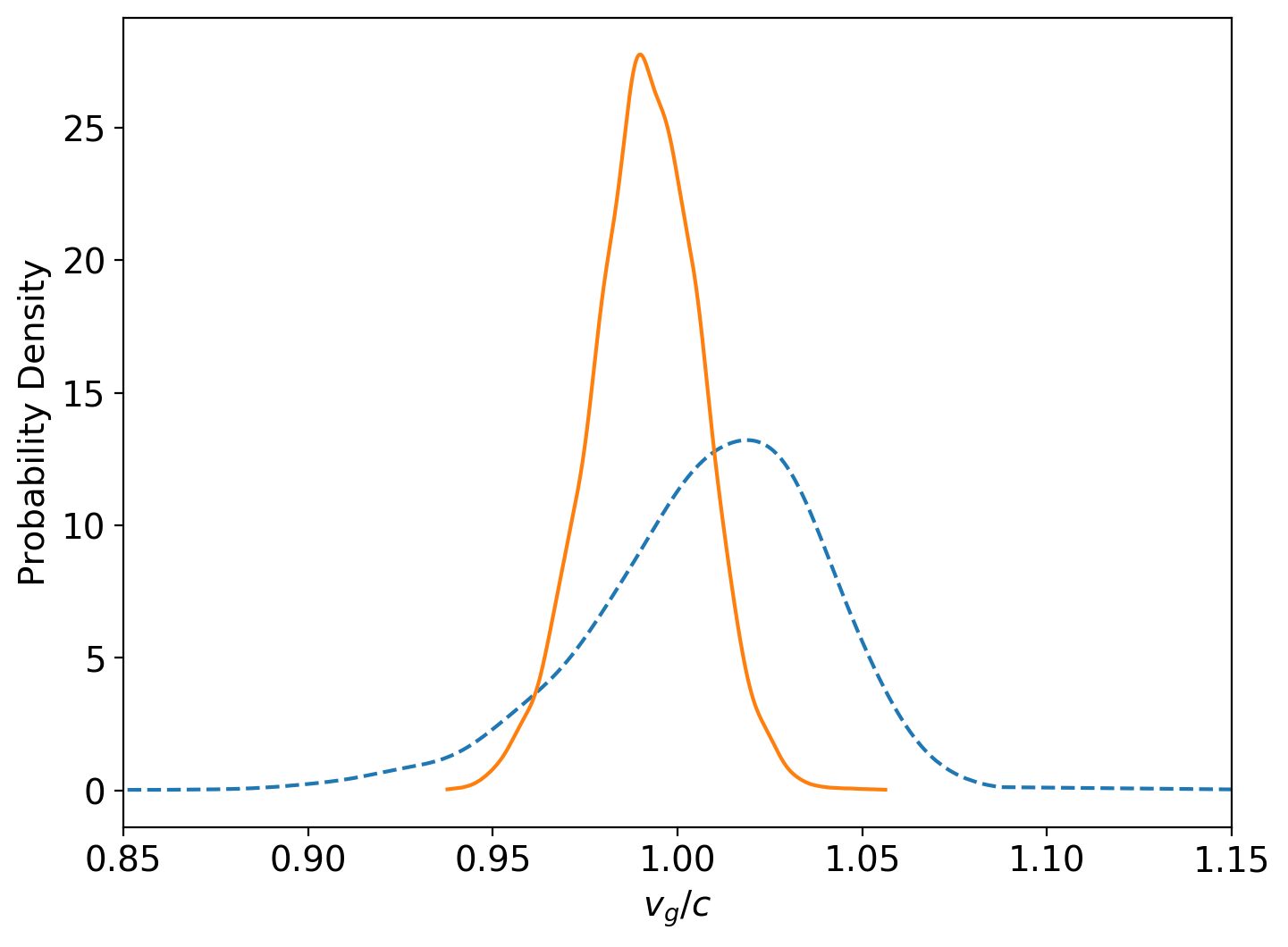}
\caption{\label {fig:gw170817} Marginalized posterior distributions of $v_g$ for GW170817. The solid line is obtained from the run with fixing $\alpha$ and $\delta$ at the electromagnetic counterpart, whereas the dashed line obtained from the run without fixing $\alpha$ and $\delta$.}
\end{figure}
 
\begin{figure}
\centering
\includegraphics[width=0.5\textwidth]{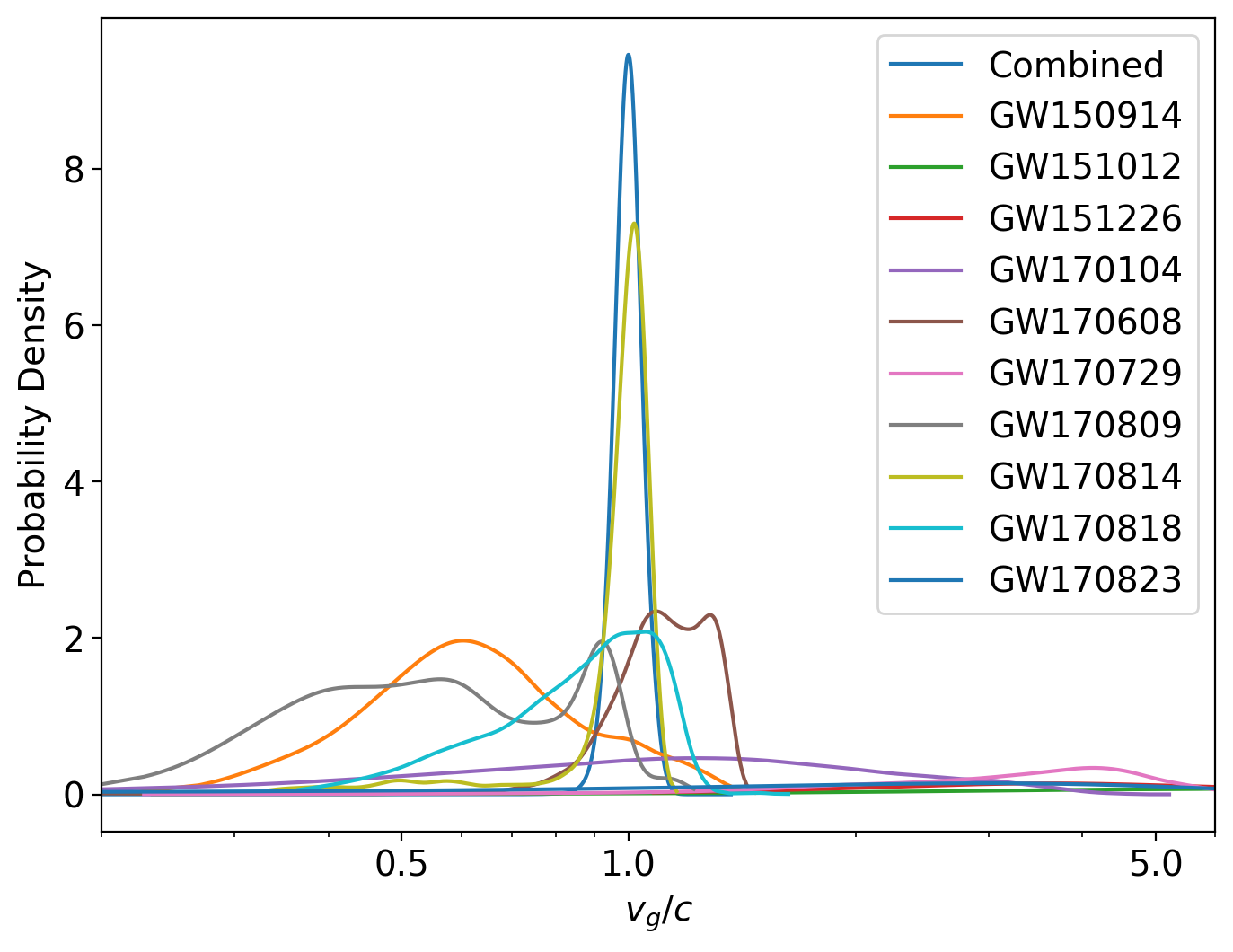}
\caption{\label {fig:BBH} Posterior distributions of $v_g$ for ten BBH events: GW150914, GW151012, GW151226, GW170104, GW170608, GW170729, GW170809, GW170814, GW170818, GW170823, and combined posterior. The combined posterior is computed using Eq. \ref{eq:simplified}.}
\end{figure}

\section{\label{sec:results} Results}

We use \textsc{\texttt{lalinference\_mcmc}} \cite{lalsuite} which implements MCMC with Metropolis-Hastings algorithm to run the Bayesian parameter estimation. In this paper, we use a uniform prior in $v_g$, the prior upper bound of $v_g$ can be estimated by using GW coalescence times \cite{FINDCHIRP} at two LIGO detectors and assuming GW source, and two LIGO detectors are on the same line. Here, we take coalescence time to be the time when the GW amplitude peaks. Suppose that for a GW event, the coalescence time at Hanford is $t_H$, Livingston is $t_L$ and distance between Hanford  and Livingston is $d$, then the prior upper bound $max(v_g) = d/(|t_H-t_L|+2\sigma)$, where $\sigma$ is the uncertainty of the coalescence time.

In our analysis, we choose the IMRPhenomPv2 waveform \cite{Hannam:2014} for all BBH events and TaylorF2 for the BNS event. IMRPhenomPv2 is a processing BBH waveform with inspiral, merger and ringdown. TaylorF2 \cite{Sathyaprakash, Vines:2011, Mikoczi, Bohe:2013, Bohe:2015, Arun:2009} is a frequency domain post-Newtonian waveform model that includes tidal effects.

The first detection of binary neutron star inspiral GW170817 by Advanced LIGO and Advanced Virgo provides an accurate measurement for $v_g$. The BNS event has a network signal to noise ratio(SNR) 33 \cite{LVC:Catalog}  which is the highest in all GW events detected in O1 and O2. The sky localization is precisely constrained to an area of 16 $deg^2$. Those two aspects of GW170817 allow an accuracy $v_g$ measurement to a $(0.95c, 1.06c)$ 90\% confidence interval. The later electromagnetic counterpart was discovered in the galaxy NGC4993 \cite{LVC:MMA}, which enable us to fix the right ascension($\alpha$) and declination($\delta$) at the electromagnetic counterpart during MCMC sampling. The later measurement shrinks the 90\% confidence interval of $v_g$ to $(0.97c, 1.02c)$. The marginalized posteriors of $v_g$ for GW170817 with and without fixing $\alpha$ and $\delta$ at the electromagnetic counterpart are shown in FIG. \ref{fig:gw170817}. 

\begin{figure}
\centering
\includegraphics[width=0.5\textwidth]{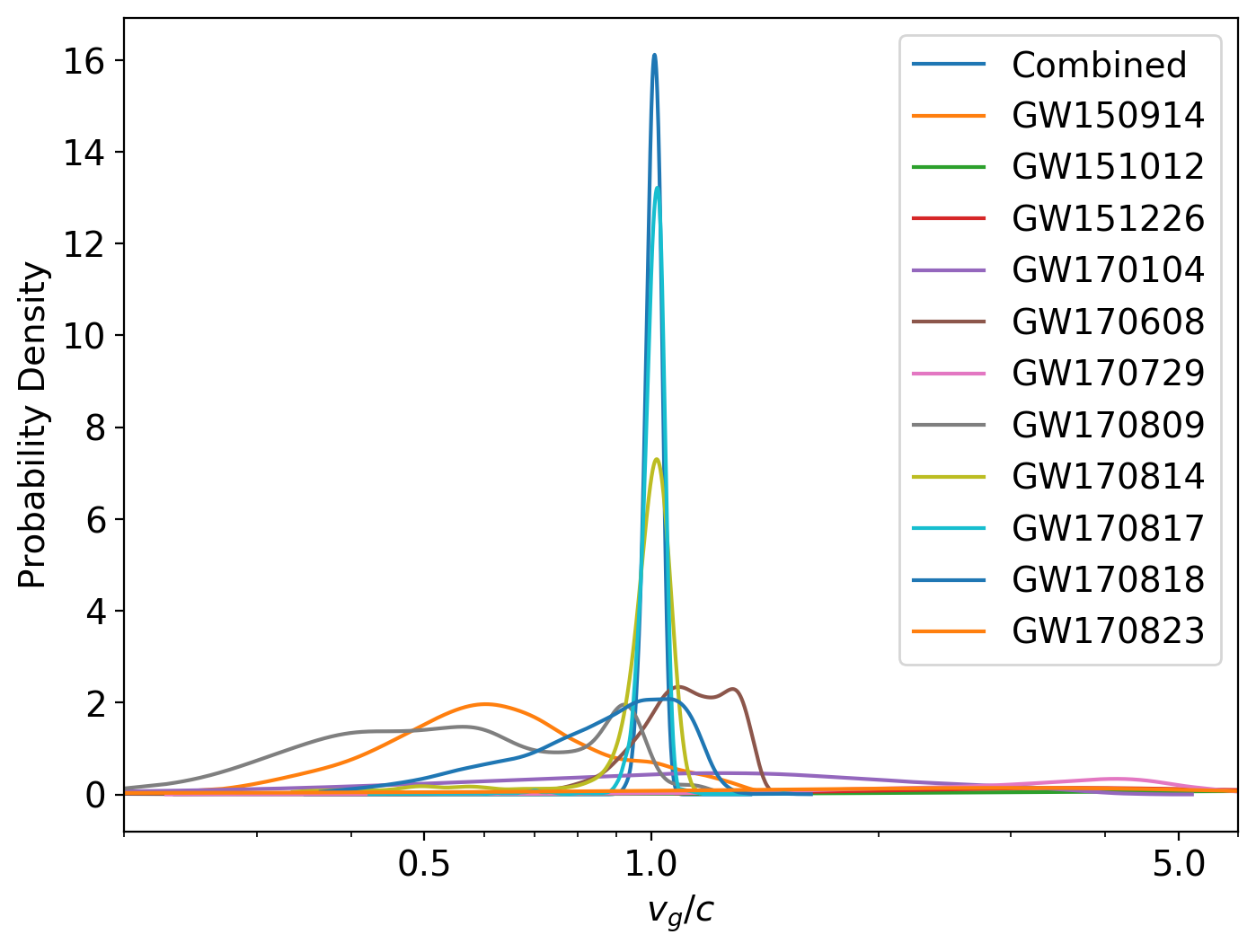}
\includegraphics[width=0.5\textwidth]{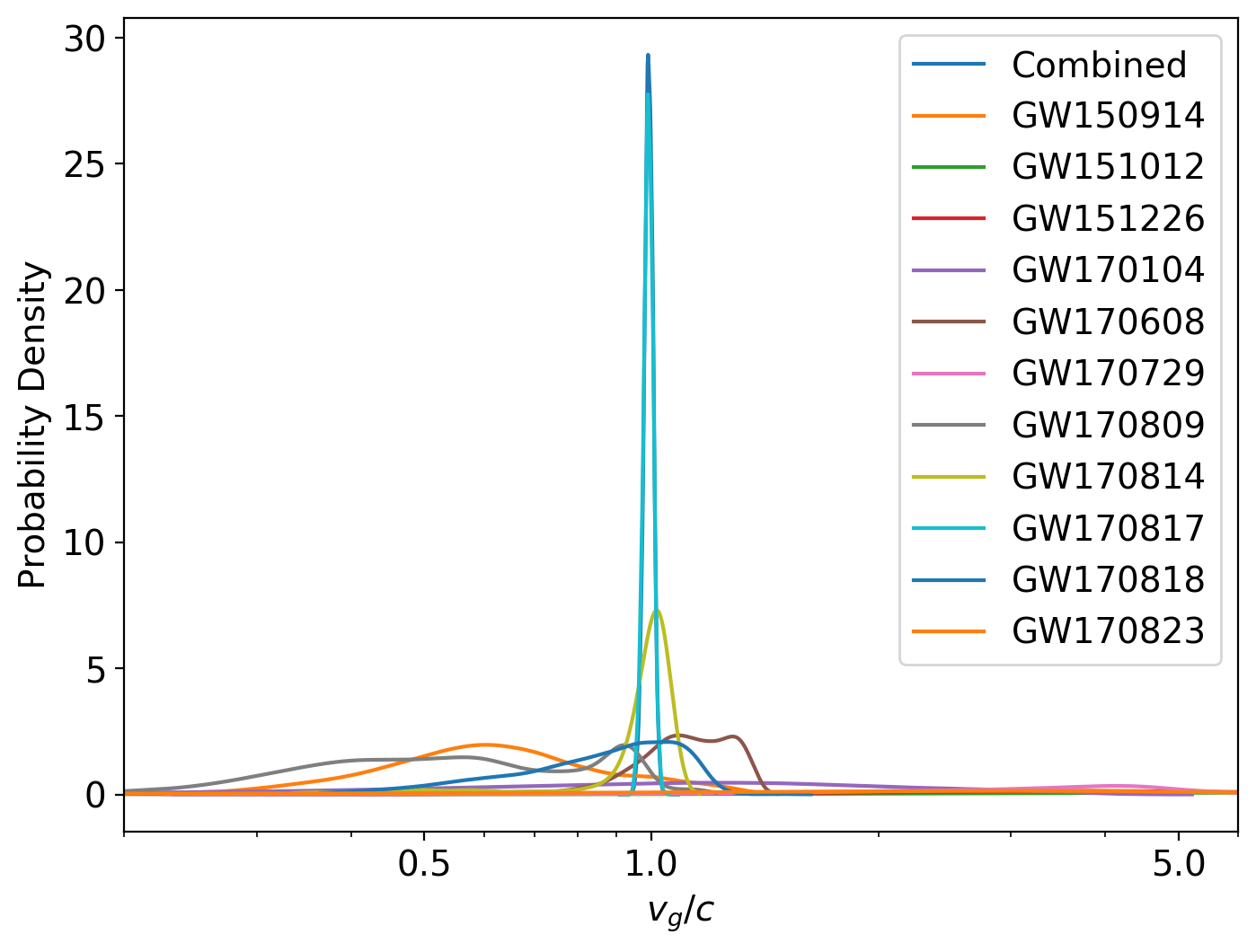}
\caption{\label {fig:combined} Posterior distributions of $v_g$ for ten BBH and a BNS detected in O1 and O2: GW150914, GW151012, GW151226, GW170104, GW170608, GW170729, GW170809, GW170814, GW170817, GW170818, GW170823, and combined posterior. For GW170817 $\alpha$ and  $\delta$ are free parameters in the top plot, in the bottom plot $\alpha$ and  $\delta$ are fixed at the electromagnetic counterpart.}
\end{figure}

\begin{table}
\caption{\label{fig:cfl} 90\% confidence intervals of $v_g$ from individual events posteriors and combined posteriors. GW170817(fixed) obtain from the MCMC run with fixing $\alpha$ and  $\delta$ at the electromagnetic counterpart and GW170817 treats $\alpha$ and  $\delta$ as free parameters. Combined(BBH) obtained from the seven BBH. Combined(fixed) and Combined uses seven BBH and GW170817 with and without fixing $\alpha$ and  $\delta$ respectively. Network SNR values are reported from the GstLAL search pipeline \cite{LVC:Catalog}. 90\% confidence regions of the sky localization ($\Omega$) with fixing $v_g$ at c are presented in GWTC-1\cite{LVC:Catalog} and without fixing $v_g$ are computed from posteriors of  $\alpha$ and  $\delta$.\\}

\scriptsize
  \centering
\begin{tabular*}{0.5\textwidth}{c @{\extracolsep{\fill}} ccccc }
\hline
&  \multicolumn{1}{c}{90\% Confidence} & \multicolumn{1}{c}{Network} & \multicolumn{1}{c}{$\Omega$/$deg^2$} & \\ 
Events &  Intervals & SNR &  GWTC-1 & $\Omega$/$deg^2$ \\[1ex]
\hline
GW150914 & (0.35c, 1.14c) & 24.4 & 182 & 2385\\[1ex]
GW151012 & (2.95c, 16.19c) & 10.0 & 1523 & 6607\\[1ex]
GW151226 & (1.22c, 12.00c) & 13.1 & 1033 & 6515\\[1ex]
GW170104 & (0.34c, 3.27c) & 13.0 & 924 & 5313\\[1ex]
GW170608 & (0.91c, 1.38c) & 14.9 & 396 & 1269\\[1ex]
GW170729 & (1.56c, 5.83c) & 10.8 & 1033 & 1287\\[1ex]
GW170809 & (0.30c, 1.01c) & 12.4 & 340 & 2252\\[1ex]
GW170814 & (0.88c, 1.11c) & 15.9 & 87 & 250\\[1ex]
GW170817 & (0.95c, 1.06c) & 33.0 & 16 & 53\\[1ex]
GW170817(fixed) & (0.97c, 1.02c) & 33.0 & 0 & 0\\[1ex]
GW170818 & (0.59c, 1.21c) & 11.3 & 39 & 168\\[1ex]
GW170823 & (0.10c, 12.19c) & 11.5 & 1651 & 6412\\[1ex]
\hline
Combined(BBH) & (0.93c, 1.07c) & &\\[1ex]
Combined & (0.97c, 1.05c) & &\\[1ex]
Combined(fixed) & (0.97c, 1.01c) & &\\[1ex]
\hline
\end{tabular*}
\label{table:clf}
\end{table}

FIG. \ref{fig:BBH} shows the posterior distributions of $v_g$ for ten O1 and O2 BBH events, and the combined posterior is obtained by using Eq. \ref{eq:simplified}. Narrow sky localization and high SNR of a GW event can help to better constrain on $v_g$. $v_g$ for GW170809, GW170814 and GW170818 are well constrained due to the fact that they were observed by the three GW detectors which can help to better localize the GW sources. GW170729 was also observed by the three detectors, due to its lower SNR, $v_g$ of GW170729 is poorly measured. GW150914 and GW170608 were only observed by the two LIGO detectors, however, due to its higher SNR, they are better constrained than GW170729. GW151012, GW151226, GW170104, and GW170823 were also observed by the two LIGO detectors, but the sky localization of these events are poorly constrained, hence posteriors of $v_g$ for these two events are relatively flat. 90\% confidence interval of $v_g$ for GW170814 is (0.88c, 1.11c) which is the best measurement among the BBH events detected in O1 and O2. The 90\% confidence interval of the combined posterior of all BBH shrinks to (0.93c, 1.07c) which improved by 30\% relative to GW170814. 90\% confidence for all individual events and the combined posteriors are listed in TABLE \ref{table:clf}.  

By including the ten BBH events and the BNS event, the combined posterior is constrained to (0.97c, 1.05c) and (0.97c, 1.01c) for GW170817 with and without fixing $\alpha$ and  $\delta$ at the electromagnetic counterpart respectively. Top plot of FIG. \ref{fig:combined} shows the combined posterior alone with posteriors of seven BBH events and GW170817 without fixing $\alpha$ and  $\delta$, and the bottom plot shows the results with fixing $\alpha$ and  $\delta$ at the electromagnetic counterpart. We can see that most of the contribution to the combined posteriors comes from GW170817 because it is measured more accuracy than other BBH events. The narrow posterior of GW170817 can also help to remove tails from the combined posteriors, the two combined posteriors with GW170817 show fewer tails than the combined posterior with BBH only.

\begin{figure*}
\centering
\includegraphics[width=0.8\textwidth]{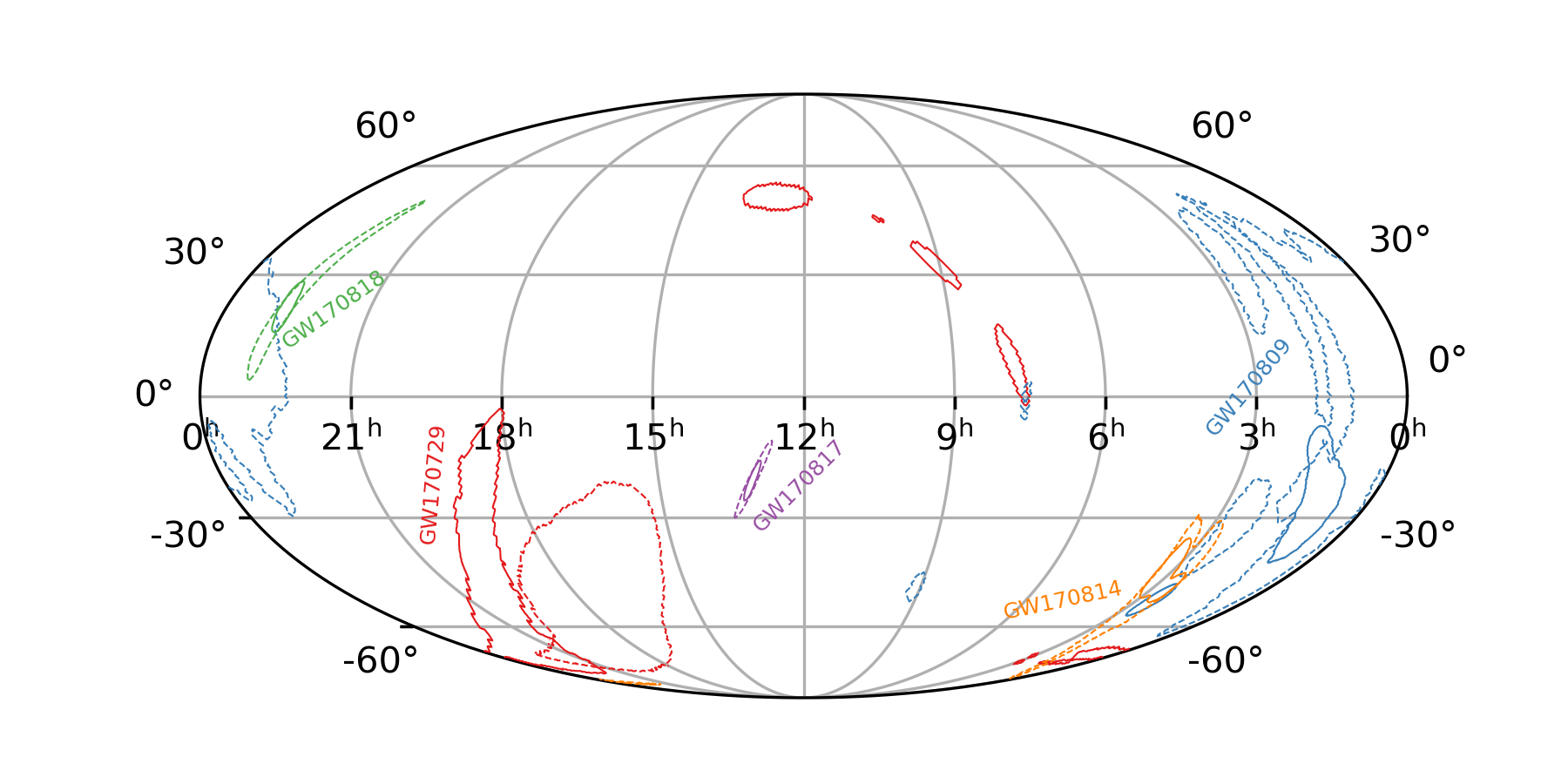}

\includegraphics[width=0.8\textwidth]{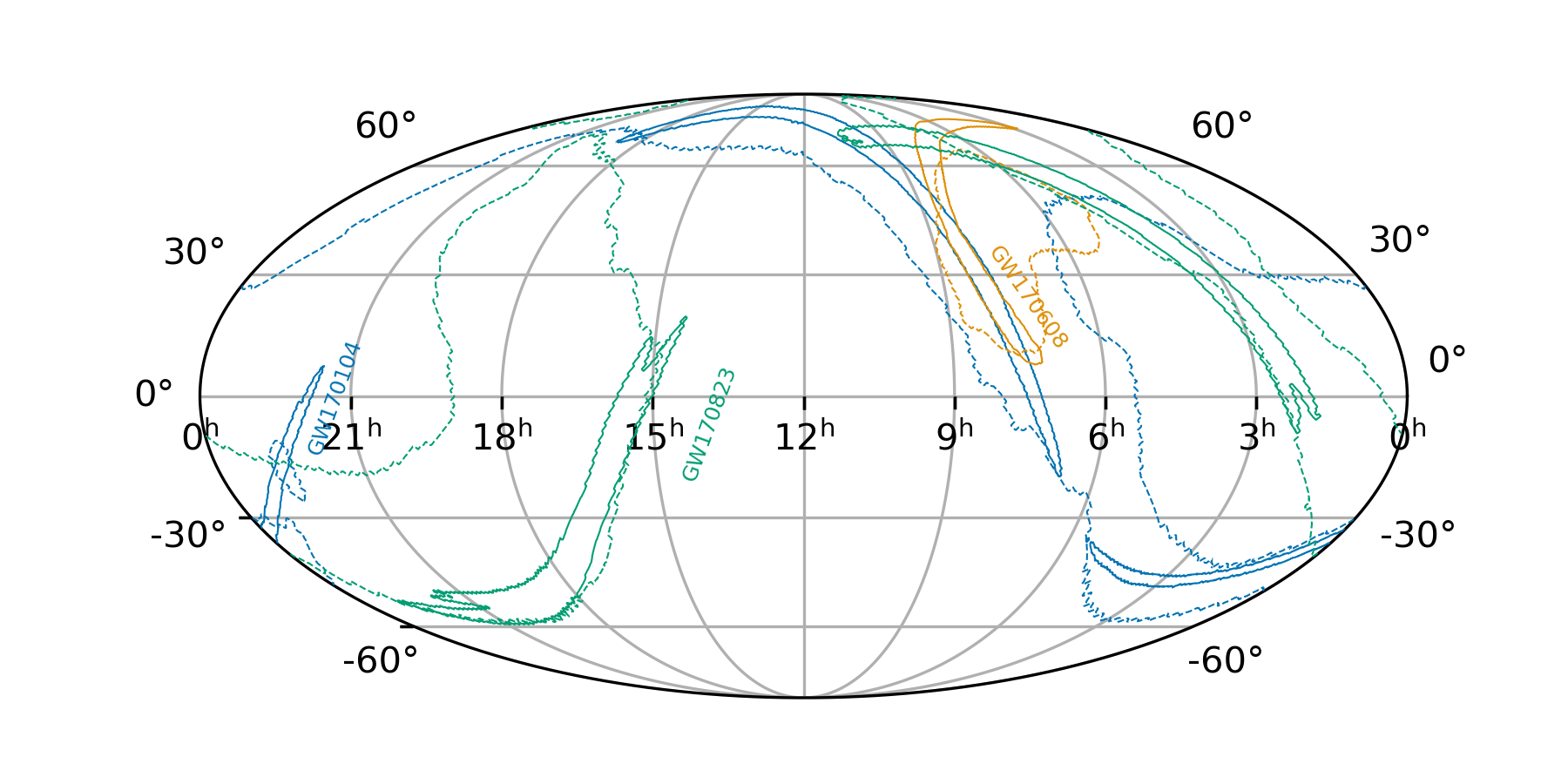}

\includegraphics[width=0.8\textwidth]{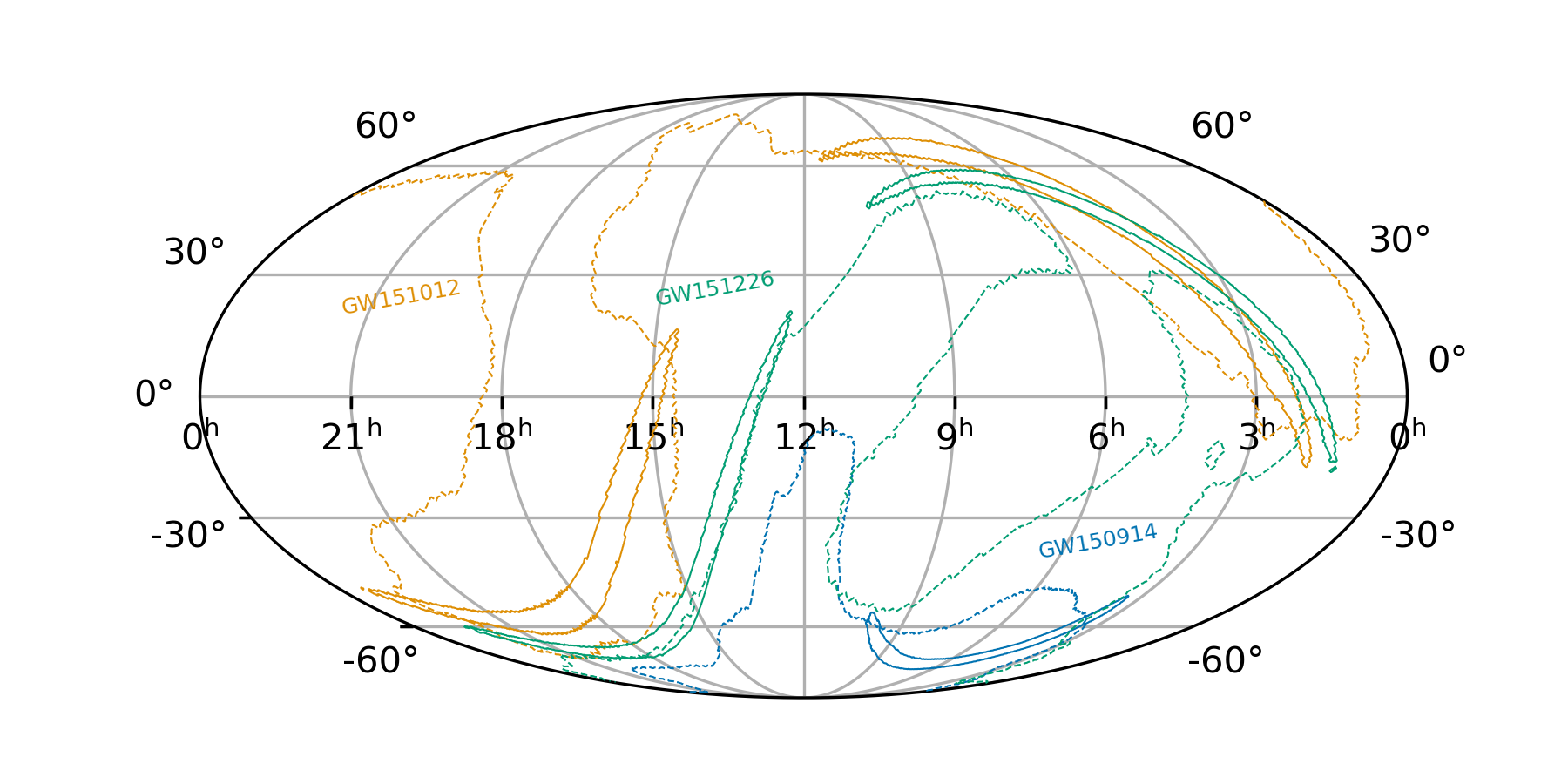}

\caption{\label {fig:skymap} 90\% confidence regions for the sky localizations of all GW events detected in O1 and O2. The solid contours are obtained from the posteriors where the speed of gravitational waves is fixed at the speed of light, the dashed contours shows the results for $v_g$ as a free parameter. Top: events detected by Advanced LIGO and Advanced Virgo (GW170729, GW170809, GW170814, GW170817, GW170818); middle: events detected by the two Advanced LIGO detectors in O2 (GW170104, GW170608, GW170823); bottom: events detected in O1 (GW150914, GW151012, GW151226).}
\end{figure*}

The speed of gravitational waves is correlated with the sky localization of a GW source. When $v_g$ is allowed as a free parameter in the parameter estimation, the uncertainty of $\alpha$ and $\delta$ tend to increase. FIG. \ref{fig:skymap} shows the comparison of skymaps with and without fixing $v_g$ at c, and the corresponding 90\% confidence regions of the sky localization are listed in TABLE \ref{table:clf}. If gravitational waves propagate at a speed different at the speed of light, the skymap obtained from the parameter estimation where $v_g$ is fixed at the speed of light could be biased.

\section{\label{sec:lorentz} Local Lorentz Violation}

The 9 nondispersive, nonbirefringent coefficients for Lorentz violation in the gravity sector of the SME cause modification of the group velocity of GWs. 
Using natural units and the assumption that the nongravitational sectors, including the photon sector, are Lorentz invariant,
the modified group velocity can be written as follows \cite{kmgw}:
\begin{equation}
v_g
= 1 + \half \sum_{jm} (-1)^j Y_{jm} (\alpha,\delta) \sb_{jm}.
\label{eq:sme}
\end{equation}
Here a basis of spherical harmonics $Y_{jm}$ 
in which $j\leq2$ has been used
to express the 9 Lorentz-violating degrees of freedom $\sb_{jm}$ present in this limit
of the SME.
While the sum on $m$ ranges from $\pm j$ in Eq.\ \rf{eq:sme},
the equivalent expansion over positive $m$:
\bea
\nonumber
v_g
= 1 + 
\sum_j (-1)^j \Big( \half \sb_{j0} Y_{j0} 
  + \sum_{m>0} && [\re \sb_{jm} \re Y_{jm}\\
  &&-  \im \sb_{jm} \im Y_{jm}]\Big), \phantom{222}
\label{eq:sme2}
\eea
is conventionally chosen in expressing experimental sensitivities.

With 11 GW events detected in O1 and O2, it is possible to simultaneously constrain all 9 of the $\sb_{jm}$ coefficients for Lorentz violation. However,
some of these have significant uncertainty in both $v_g$ and sky position $\alpha,\delta$.
Hence we explore a model formed by the $j\leq1$ subspace of the full SME,
using 4 of the most sensitive events
and the following methods.

In the earlier sections of the paper,
data from the multiple events were combined under the assumption of isotropic GW speeds
to obtain a more sensitive measurement.
Here we exploit a complementary advantage of the multiple observations
in constraining direction-dependent speeds.
To develop the methods,
imagine that one had an exact measurement of $v_g$ as well as sky position for a GW event.
Then Eq.\ \rf{eq:sme2} would form 1 equation with 4 unknowns (the 4 coefficients $\sb_{jm}$ in our model).
Given 4 such events, assuming unique sky locations, 
the system of 4 equations that results could be solved for the 4 coefficients $\sb_{jm}$
forming a measurement of Lorentz violation.
Of course in the present case of real experimental work 
we have a distribution for each of our 4 events rather than a signal value.
We use this data by randomly drawing a sample from the distribution for each of the 4 events,
solving for the corresponding values of the 4 coefficients $\sb_{jm}$,
and repeating the process to build the $\sb_{jm}$ distribution.

\begin{figure}
\centering
\includegraphics[width=0.5\textwidth]{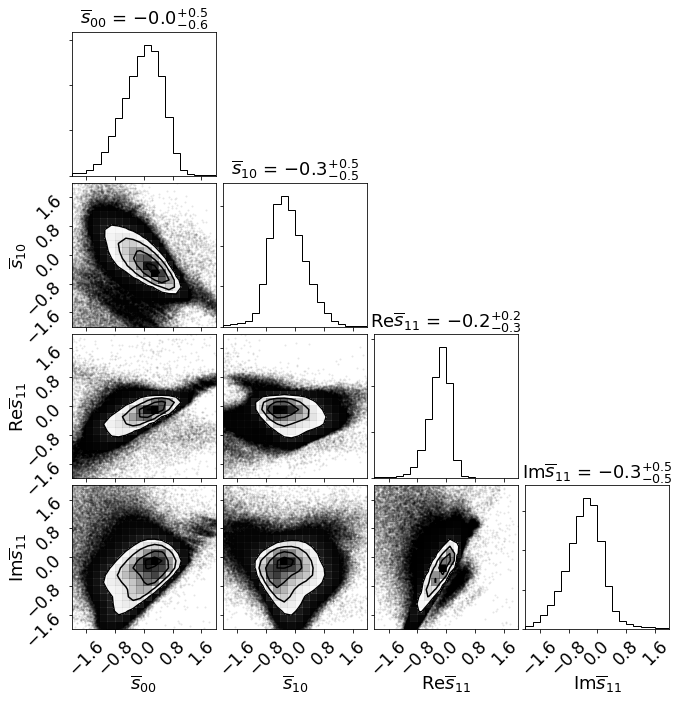}

\caption{\label {fig:lv1} The distribution of $\sb_{jm}$ values implied by 
the events listed on lines 2, 5, 6, and 8 of Table \ref{fig:cfl}.
Numbers above the plots show best values
with a one sigma range.}
\end{figure}

\begin{figure}
\centering
\includegraphics[width=0.5\textwidth]{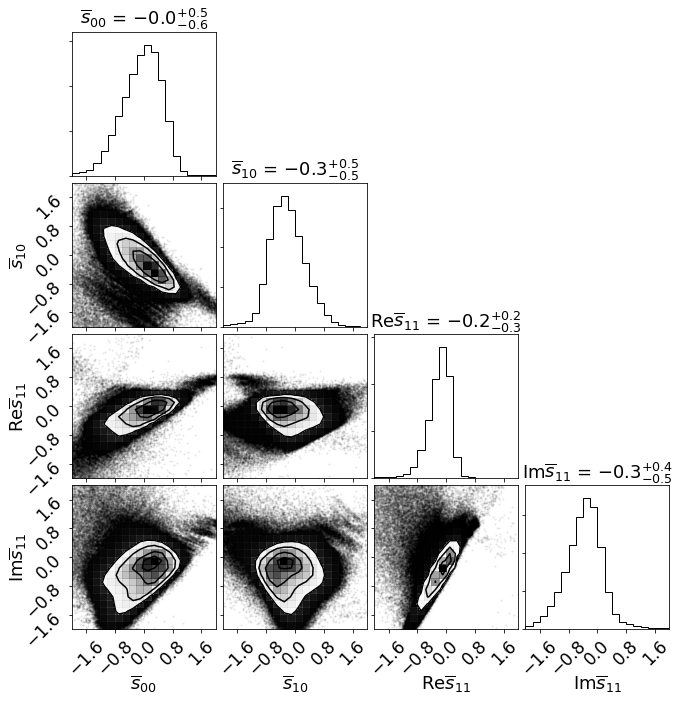}

\caption{\label {fig:lv2} The distribution of $\sb_{jm}$ values implied by 
the events listed on lines 2, 5, 7, and 8 of Table \ref{fig:cfl}.
That is,
relative to Fig.\ \ref{fig:lv1},
the sky position of GW170817 is fixed.
Numbers above the plots show best values
with a one sigma range.}
\end{figure}

Using GW170608, GW170814, GW170817, and GW170818 (lines 2, 5, 6, and 8 of Table \ref{fig:cfl})
we obtain the results shown in Fig.\ \ref{fig:lv1}.
In Fig.\ \ref{fig:lv2},
we use the same events but with the fixed sky position as in line 7 of 
Table \ref{fig:cfl}.
This generates a modest narrowing of the one sigma range for some coefficients.
This generates only small changes in the plot. We also explored setting the speed of gravity for the GW170817 event to that found in Ref.\ \cite{LVC:2017gw-grb}.
This also results in an insignificant effect on the confidence bands.

Note that the measurements of the $\sb_{jk}$ shown in Figs.\  \ref{fig:lv1} and  \ref{fig:lv2}
are consistent with zero.
Hence we can interpret the one sigma range shown as upper and lower bounds on the values of the 
$\sb_{jk}$ coefficients,
an exclusion of the simplest types of direction-dependent speeds.
As with $v_g$,
these limits are considerably weaker than some found in the literature \cite{data}.
However,
they carry value in that they are obtained from significantly different methods than other tests,
are the first effort to simultaneously constrain multiple $\sb_{jk}$ using speed of gravity 
measurements,
and begin establishing methods for future higher-sensitivity tests.

As a final note,
we point out that if the isotropic limit of the SME is considered
such that $\sb_{00}$ is the only nonzero coefficient for Lorentz violation,
then the combined $v_g$ results in the last line of Table \ref{fig:cfl}
may be applied.
Doing so yields $-0.2 < \sb_{00} < 0.07$.

\section{Conclusions}
While the association between GWs and gamma-rays observed with GW170817 and GRB 170817A have provided an extremely tight bound on the difference between the speed of gravitational waves and the speed of light, in this paper we have presented an independent method of directly measuring $v_g$, which, while less precise, is based solely on GW observations and so not reliant on multimessenger observations.  We continue to find measured values of $v_g$ consistent with the speed of light, as predicted by General Relativity, not just for GW170817 but also for other signals detected during the second observation run of Advance LIGO and Virgo.  By combining these measurements and assuming isotropic propagation, we constrain the speed of gravitational waves to (0.97c, 1.01c) which is within 3\% of the speed of light in a vacuum.
We also obtain simultaneous constraints on nonbirefringent, nondispersive 
coefficients for Lorentz violation in the test framework of the SME.
Though the constraints are not as strong as other methods,
we simultaneously limit multiple coefficients using direct speed of gravity tests
for the first time,
directly constraining the possibility of an anisotropic speed of gravity. Other implications for deviations from general relativity arising from cosmological evolution were considered in Ref. \cite{2020JCAP}.

There are some limitations of the approach used here that must be acknowledged.  First, should the speed of gravitational waves differ from the speed of light — in violation of the predictions of General Relativity — then we would not necessarily expect other assumptions based on General Relativity predictions to necessarily hold.  Among the assumptions that could affect us is the assumption that the gravitational waves only exist in two tensorial transverse polarizations. More general metric theories of gravity could allow for up to 6 independent polarizations, including in addition two longitudinal-vector polarizations and two scalar polarizations (one longitudinal and one transverse, thought these are indistinguishable in interferometric detectors). The observations of GW170817 as well as others events detected in O1 and O2 show that the pure tensor mode is strongly favored over the pure vector or pure scalar mode \cite{{LVC:GW170817TGR}, {catalogTGR}}. The dominant tensor mode could be mixed with smaller scalar and/or vector modes propagating with the same speed. Even in that case, the presence of the vector mode has already constrained weakly with GW170817 by constructing a null stream \cite{2019PhRvD.100f4010H}. Our parameter estimation has continued to assume that only the tensor polarization states exists.  Furthermore, we continue to assume that the gravitational waveforms are as predicted by general relativity.  Nevertheless, our measurement of $v_g$ is mostly constrained by the measured times of arrival of the signal in the various detectors, so we believe it is reasonably robust.

In addition, as noted in Ref. \cite{Cornish:2017sgw} the searches that identify GW signals normally require a signal to be seen in two detectors and impose a time window.  For example, for the LIGO Hanford Observatory and the LIGO Livingston Observatory, the searches required arrival times within 15 ms (while the light travel time between those detectors is 10 ms) \cite{LVC:GW150914S}.  This would seemingly create a selection bias against gravitational wave signals with $v_g < \frac{2}{3}c$.  However (again as noted in Ref. \cite{Cornish:2017sgw}) a gravitational wave signal can also be identified in a single detector, and, for sufficiently loud signals, the presence of a signal in another detector at similar time would unlikely go unnoticed.  For this reason we do not think there is a strong selection bias against slow moving gravitational waves.

\section{\label{sec:level1}ACKNOWLEDGMENTS}
The authors would like to thank Patrick Brady, Will Farr, Ignacio Magana Hernandez, and Eric Thrane for useful discussions. The authors would like to thank Leila Haegel for useful feedback during the internal LIGO-Virgo Collaboration review. We thank LIGO and Virgo Collaboration for providing the data from the second observing run. This work was supported by NSF awards PHY-1607585, PHY-1912649, and PHY-1806990. The authors are grateful for computational resources provided by the LIGO Laboratory and supported by National Science Foundation Grants PHY-0757058 and PHY-0823459.

\bibliography{Speed_of_Gravity}

\end{document}